\documentclass{article}

\usepackage{amsmath, amsthm, amsfonts}

\title{Central exclusive production in the ALICE experiment at the LHC}
\author{R. Schicker\\
  \small Physikalisches Institut, Heidelberg University,\\
  \small Im Neuenheimer Feld 226, 69120 Heidelberg, Germany\\
  \small schicker@physi.uni-heidelberg.de
}

\begin{document}
\maketitle

\abstract{The ALICE experiment at the Large Hadron Collider (LHC) at CERN consists of a
central barrel, a muon spectrometer and additional detectors for trigger and event 
classification purposes. The low transverse momentum threshold of the central barrel gives
ALICE a unique opportunity to study the low mass sector of central exclusive production
at the LHC.

Keywords: ALICE experiment; exclusive production; gluon shadowing.
}

\section{Introduction}

The ALICE experiment at CERN consists of a central barrel and of a forward muon
spectrometer\cite{Alice1}. Additional detectors for trigger purposes and event classification
exist outside of the central barrel. Such a geometry allows the investigation of many
properties of diffractive reactions at hadron colliders, for example the measurement
of single and double diffractive dissociation cross-sections and the study of central
exclusive production (CEP). The ALICE experiment has taken data in proton-proton, 
proton-lead as well as lead-lead collisions in Run I at the Large Hadron Collider (LHC).

\section{The ALICE Experiment}

In the ALICE central barrel, momentum reconstruction and particle identification
are achieved in the pseudorapidity range $ -1.4 < \eta < 1.4$  by com- bining the information 
of the Inner Tracking System (ITS) and the Time Projection Chamber (TPC). In the 
pseudorapidity range \mbox{$-0.9 < \eta < 0.9$}, the information from the
Transition Radiation Detector (TRD) and the Time-of-Flight (TOF) system is also
available. A muon spectrometer covers the range $ -4.0 < \eta < -2.5$. At very forward 
angles, the energy flow is measured by Zero Degree Calorimeters (ZDC)\cite{ZDC}.
Detectors for event classification and trigger purposes are located on both sides
of the ALICE central barrel. First, the scintillator arrays V0A and V0C cover the
pseudorapidity range $ 2.8 < \eta < 5.1$ and $-3.7 < \eta < -1.7$, respectively. The 
four- and eight-fold segmentation in pseudorapidity and azimuth result in 32 individual
counters in each array. Second, a Forward Multiplicity Detector (FMD) based on
silicon strip technology covers the pseudorapidity range $-3.4 < \eta < -1.7$ and
$1.7 < \eta < 5.1$, respectively. Additional scintillator arrays ADA and ADC are being
built for data taking in \mbox{Run II}, with pseudorapidity coverage of $ -7.0 < \eta < -4.8 $
and $4.7 < \eta < 6.3$, respectively.

\section{Central Production in ALICE}

Central diffractive events are experimentally defined by activity in the central barrel
and by no activity outside the central barrel. This condition can be implemented
at trigger level zero (L0) by defining barrel activity as hits in the ITS pixel 
detector (SPD) or the TOF system. The gap condition is realized by the absence of
V0 signals, hence a gap of two units in pseudorapidity on either barrel side can be
defined at L0. In Run II, the additional detectors ADA and ADC will further enhance 
the L0 gap trigger capabilities as decribed above. In the offline analysis, the
information from the V0, FMD, ITS and TPC detectors define the gaps spanning
the range $ -3.7 < \eta < -0.9$  and $ 0.9 < \eta < 5.1$. Events with and without detector 
signals in these two ranges are defined to be no-gap and double-gap events,
respectively. A rapidity gap can be due either to Pomeron, Reggeon or photon exchange. 
A double-gap signature can therefore be induced by a combination of these
exchanges. Pomeron–Pomeron events result in centrally produced states with quantum 
numbers C = +1 (C = C-parity) and I = 0 (I = isospin). The corresponding
quantum numbers in photon-Pomeron induced events are C = -1 and \mbox{I = 0} or
I = 1\cite{Nachtmann}. Exclusive particle production in proton-proton collisions is 
dominated by Pomeron-Pomeron interactions, whereas photon-photon and Pomeron-photon 
exchanges dominate in lead-lead collisions. The systematics of Pomeron and photon
interactions can therefore be studied by analyzing proton-proton, proton-lead and
lead-lead collisions.

\section{Central Meson Production in Proton-Proton Collisions}

In the years 2010-2011, ALICE recorded zero bias and minimum bias data in
pp-collisions at a center-of-mass energy of $\sqrt{s}$ = 7 TeV. The zero bias trigger
was defined by beam bunches crossing at the ALICE interaction point, while the
minimum bias trigger was derived by minimum activity in either the ITS pixel or
the V0 detector. Events with double-gap topology as described above are contained
in this minimum bias trigger, hence central diffractive events were analyzed from
the minimum bias data sample. For the results discussed below, 3.6$\times$10$^8$ minimum
bias events were analyzed. First, the fraction of events satisfying the gap condition
described above was calculated. This fraction was found to be about 2$\times$10$^{-4}$ .
Only runs where this fraction was calculated to be within 3$\sigma$ of the average value
of the corresponding distribution were further analyzed. This procedure resulted
in about 7$\times$10$^4$ double-gap events. As a next step, the track multiplicity in the
pseudorapidity range $-0.9 < \eta < 0.9$ was evaluated. Very low transverse momentum
tracks never reach the TPC which results in events with track multiplicity zero. The
multiplicity distributions of the double- and no-gap events clearly show different
behavior as discussed in Ref. \cite{Schicker}.

The specific energy loss dE/dx as measured by the TPC in combination
with the TOF detector information identifies pions with transverse momenta
p$_T \ge $ 300 MeV/c. The events with exactly two pions are selected, and the invariant
mass of the pion pairs is calculated. These pion pairs can be of like- or unlike-sign
charge. Like-sign pion pairs can arise from two-pion pair production with loss of one
pion of same charge in each pair, either due to the low p$_T$-cutoff described above, or
due to the finite pseudorapidity coverage of the detectors used for defining the 
rapidity gap. For charge symmetric detector acceptances, the unlike-sign pairs contain
the signal plus background, whereas the like-sign pairs represent the background.
From the two distributions derived in the analysis, the background is estimated
to be less than 5\%. The particle identification by the TOF detector requires the
single-track transverse momentum p$_T$ to be larger than about 300 MeV/c. This
single-track p$_T$-cut introduces a significant acceptance reduction for pair masses
\mbox{M($\pi\pi$) $\le$ 0.8 GeV/c$^2$} at low pair-p$_T$. In the no-gap events, structures 
are seen from $K_s^0$ - and $\rho_0$-decays. Two additional structures are associated 
with f$_0$(980)- and f$_2$(1270)-decays. In the double-gap distribution, the $K_s^0$ and 
$\rho_0$ are highly suppressed while the f$_0$(980) and f$_2$(1270) with quantum numbers 
\mbox{$J^{PC}$ = (0,2)$^{++}$} are much enhanced\cite{Reidt}. This enhancement of 
\mbox{$J^{PC}$ = (even)$^{++}$} states is evidence that the double gap condition used for 
analysing the minimum bias data sample selects events dominated by double Pomeron exchange.

\section{Central Meson Production in Lead-Lead Collisions}

Heavy-ion and proton beams at high energies are the sources of strong electromagnetic 
fields. These fields can be represented by an ensemble of equivalent photons by identifying 
the Poynting vector of the electromagnetic field with the corresponding quantity of the 
photon ensemble\cite{Fermi,Jackson1}. Cross-sections of heavy-ion and proton-induced
photon-photon processes at high energies can subsequently be calculated by folding
the respective photon flux with the elementary photon-photon cross-section taking into 
account the electromagnetic form factor\cite{Budnev,Jackson2,Krauss,Baur,Baltz}. The 
photon flux associated with the heavy-ion electromagnetic field scales with the nuclear 
charge squared, hence large cross-sections for heavy-ion induced electromagnetic and 
photonuclear processes result. Of particular interest is exclusive photoproduction of 
vector mesons in heavy-ion collisions. The vector meson production cross-section depends 
on the nuclear gluon distribution, hence allows the study of nuclear gluon shadowing 
effects at values of Bjorken-x $\sim 10^{-3}$ and $\sim 10^{-2}$ for data taken in the ALICE 
central barrel and the muon spectrometer, respectively.

During the heavy-ion runs in 2010 and 2011, special triggers were implemented
to filter out gap topologies. In the 2011 run, a trigger to select forward produced
J/$\psi$ consisted of a single muon trigger in the muon spectrometer, of at least one
hit in the V0C detector, and of no hit in the V0A detector. The one hit in V0C is
required due to the large overlap of V0C with the muon spectrometer acceptance.
With this trigger, a sample of 3.16$\times$10$^6$ events was collected. The analysis of 
this data sample with further requirements such as track matching between muon trigger
chambers and tracking chambers, of two reconstructed muons of opposite charge, and of 
restricting the di-muon rapidity to the range \mbox{$-3.6 < y < -2.6$} reduces the
data sample to 3209 events which prominently show the J/$\psi$ mass peak in the
invariant mass spectrum\cite{jpsi_forw}. The J/$\psi$ cross-section derived in this 
ALICE analysis is best produced by models which include nuclear gluon shadowing 
consistent with the EPS09 or EPS08 parametrizations.

In addition, a trigger to select J/$\psi$ at mid-rapidity was implemented. Activity
in the central barrel was required by at least two hits in the SPD of the ITS, a number
of TOF pads $N^{TOF}$ with signal $2 \le N^{TOF} \le 6$, with at least two of them with a 
difference in azimuth $\Delta\phi$ in the range \mbox{$150^0 \le \Delta\phi \le 180^0$}. 
The back-to-back condition applied to the TOF signals effectively restricts the final state
invariant mass to values \mbox{$\ge$ 2 GeV/c$^2$}. The absence of V0 signals is required for
the gap condition. With this trigger, a data sample of about 6.5$\times$10$^6$ was taken.
The analysis of this data sample with further requirements such as reconstructed
primary vertex, only two good tracks with tighter quality cuts with at least one of
them with $p_T \ge$ 1 GeV/c, and two track invariant mass in the range 
$2.2 < M_{inv} < 6$ GeV/c$^2$ reduces the data sample to 4542 events. The dE/dx distribution 
derived from the TPC information clearly separates the muon pairs from the electron pairs.
The background contained in the data can be estimated from the like-sign pair
distribution. The unlike-sign pair mass distribution clearly shows the J/$\psi$ mass
peak in both the electron and the muon decay channel. The cross-section derived
in this analysis is found to be in good agreement with the model which includes
the EPS09 parametrization of nuclear gluon shadowing\cite{jpsi_cent}.

\section{Conclusions}

The ALICE experiment has analysed exclusive particle production in proton-proton
and lead-lead collisions in Run I of the LHC. The comparison of properties of
double-gap events with no-gap events clearly demonstrates the justification of such
an approach. Further studies are under way for identifying the remaining background, 
and for establishing algorithms for further background reduction. Dedicated
triggers implemented in the heavy-ion run combine the special characteristics of the
exclusively produced tracks with the gap topology of the event. The analysis of these
events establishes the identification of J/$\psi$-decays, both in the ALICE central barrel 
and in the forward muon spectrometer. New detector arrays ADA and ADC will
increase the pseudorapidity coverage considerably in Run II.

\section*{Acknowledgments}

This work is supported by the German Federal Ministry of Education and 
Research under promotional reference 06HD197D and by WP8 of the hadron 
physics programme of the 7th and 8th EU programme period.



\begin{thebibliography}{99}

\bibitem{Alice1} ALICE Collab. (K. Aamodt et al.), JINST 3 (2008) S08002.

\bibitem{ZDC} R. Arnaldi et al., Nucl. Instrum. Methods A 564, 235 (2006).

\bibitem{Nachtmann} O. Nachtmann, Annals of Physics, 209 436 (1991), 
and references therein.

\bibitem{Schicker} For the ALICE Collab. (R. Schicker), \mbox{Central Diffraction in ALICE}, 
in {\it Proc. 14$^{th}$ Workshop on Elastic and Diffractive Scattering,}
Qui Nhon, 2011, eds. M. A. Rotondo and Ch. Tan (SLAC-eConf-C111215).

\bibitem{Reidt} For the ALICE Collab. (F. Reidt), AIP Conf.Proc. 1523 17 (2012).     

\bibitem{Fermi} E. Fermi, Z.Phys. A29 315 (1924), doi:10.1007/BF03184853.

\bibitem{Jackson1} J.D. Jackson, Classical Electrodynamics, 
(John Wiley \& Sons, 1962).

\bibitem{Budnev} V.M. Budnev, I.F. Ginzburg, G.V. Meledin and V.G. Serbo,
Phys.Rep.15 181 (1975).  

\bibitem{Jackson2} R.N. Cahn and J.D. Jackson, Phys.Rev.D42, 3690 (1990). 

\bibitem{Krauss} F. Krauss, M. Greiner and G. Soff, Prog.Part.Nucl.Phys., 
39, 503 (1997).

\bibitem{Baur} G. Baur, K. Hencken, D. Trautmann, S. Sadovsky and Y. Kharlov,
Phys.Rep.364, 359 (2002). 

\bibitem{Baltz} A.J. Baltz et al., Phys.Rep.458, 1 (2008). 

\bibitem{jpsi_forw} ALICE Collab. (B. Abelev et al.) 
Phys.Lett.B718, 1273 (2013). 

\bibitem{jpsi_cent} ALICE Collab. (E. Abbas et al.), 
Eur.Phys.J.C73 2617 (2013). 



\end{thebibliography}
\end{document}